\documentstyle[aps,tighten,epsf,preprint,floats]{revtex}

\def\bi{\bibitem}

\newcommand{\spartial}{\hspace{-1.2mm}\not\hspace{-.7mm}\partial}
\newcommand{\pslash}{\not\hspace{-1.mm}p}

\newcommand{\dG}{\delta {G}}
\newcommand{\dK}{\delta {K}}
\newcommand{\GG}{G}
\newcommand{\K}{K}
\newcommand{\be}{\begin{equation}}
\newcommand{\ee}{\end{equation}}
\newcommand{\eqn}[1]{\label{#1}}
\newcommand{\eq}[1]{Eq.~(\ref{#1})}
\newcommand{\eqs}[1]{Eqs.~(\ref{#1})}
\newcommand{\fign}[1]{\label{#1}}
\newcommand{\fig}[1]{Fig.~\ref{#1}}
\newcommand{\bpsi}{\bar{\psi}}

\newcommand{\AmS}{{\protect\the\textfont2
  A\kern-.1667em\lower.5ex\hbox{M}\kern-.125emS}}

\title{Complete set of electromagnetic corrections to the nucleon mass in the
Nambu-Jona-Lasinio model}

\author{A. N. Kvinikhidze\thanks{On leave from The Mathematical Institute of
Georgian Academy of Sciences, Tbilisi, Georgia.}
        and
        B. Blankleider\address{Department of Physics, The Flinders University
        of South Australia, \\
        Bedford Park, SA 5042, Australia}\thanks{The authors would like to
thank the Australian Research Council for their financial support.}}
\begin{document}
\maketitle

\begin{abstract}
We show how to derive the complete set of electromagnetic corrections to the
Nambu-Jona-Lasinio (NJL) model of the nucleon. Our results enable an accurate
estimate of the electromagnetic contribution to the neutron-proton mass
difference within this model. At the same time, our procedure demonstrates the
way to calculate the complete set of meson corrections to the NJL model that
maintains chiral symmetry.
\end{abstract}

\section{INTRODUCTION}

We have recently shown how to calculate all possible electromagnetic
corrections, of order $e^2$, to any quark or hadronic model whose strong
interactions are described nonperturbatively by integral equations
\cite{Adel98}. Here we would like to apply our method to the three-quark NJL
model of the nucleon \cite{njl} and in this way derive the expression for the
neutron-proton mass difference that is due to a complete set of electromagnetic
interaction within this model.

For this purpose it is useful to summarise the main model-independent results
obtained in Ref.\ \cite{Adel98}. Within the framework of relativistic quantum
field theory, the strong interaction Green function $G$ describing a system of
quarks or hadrons is given nonperturbatively by the integral equation whose
symbolic form is
\be
G=G_0+G_0 K G.       \eqn{G}
\ee
The complete set of lowest order electromagnetic corrections to the Green
function $G$, denoted by $\dG$, then follows from \eq{G} on topological grounds:
\be
\dG=\dG_0+\dG_0KG+G_0\dK G+G_0K\dG
+\left(G^\mu_0 K^\nu G+
G^\mu_0K G^\nu+G_0K^\mu G^\nu\right)D_{\mu\nu}  \eqn{dG}
\ee
where $\delta G_0$ is the complete set of electromagnetic corrections to the
free Green function $G_0$, and $D_{\mu\nu}$ is the photon propagator that
connects the appropriate currents (quantities with a $\mu$ or $\nu$
superscript). Unlike $\dG$ which has internal photons coupled everywhere, $\dK$
consists of the strong interaction potential $K$ with all possible photon
insertions {\em except} those that start or finish on an external quark or
hadron leg. All currents $G^\mu$, $G_0^\mu$, and $K^\mu$ are constructed by the
gauging of equations method \cite{nnn} that effectively attaches external
photons in all possible ways to the corresponding strong interaction quantities
$G$, $G_0$, and $K$, respectively. Using this method one obtains
\be
\GG^\mu = \GG \Gamma^\mu\GG, \hspace{1cm}
\Gamma^\mu = \Gamma_0^\mu + \K^\mu,\hspace{1cm}   
\Gamma_0^\mu = \GG_0^{-1} \GG_0^\mu \GG_0^{-1}. \eqn{G^mu}
\ee
With the current $G^\mu$ specified in this way, \eq{dG} can be formally solved
to give
\be
\dG = \GG\Delta\GG,\hspace{1cm}   
\Delta= \dK + \GG_0^{-1}\dG_0\GG_0^{-1}
+\left(\Gamma^\mu\GG\Gamma^\nu-\Gamma_0^\mu \GG_0\Gamma_0^\nu\right) D_{\mu\nu}.
\eqn{Delta}
\ee
The quantity $\Delta$ as given by \eq{Delta} is the key result that describes
the complete set of electromagnetic corrections to any observable of the strong
interaction model in question. For example, if the strong interactions admit a
bound state of mass $M$ and wave function $\psi$, then the complete set of
lowest order electromagnetic corrections to $M$ is given by $\delta M =
\bpsi\Delta\psi/2M$. It is a feature of our approach that the gauge invariance
of such electromagnetic corrections is a result of their completeness.

\section{NJL MODEL FOR THE NUCLEON}
The simplest NJL Lagrangian density ${\cal L}$ is defined in terms of the
iso-doublet (two flavours) colour-triplet quark field $\Psi$ as
\be
{\cal L}=\bar\Psi\left(i\spartial-m_0\right)\Psi+G\left[(\bar\Psi\Psi)^2-
(\bar\Psi\gamma_5\mbox{\boldmath $\tau$}\Psi)^2\right]  \eqn{NJLL}
\ee
where {\boldmath $\tau$} is the vector of isospin Pauli matrices, and $m_0$ is
the bare quark mass [for $m_0=0$ \eq{NJLL} is chiral invariant]. The model of
the nucleon considered here is described by the three-quark wave function that
satisfies a four-dimensional Bethe-Salpeter (BS)-like three-body integral
equation with pair interaction kernels given by the lowest order $qq$
irreducible diagrams corresponding to the Lagrangian of \eq{NJLL}, namely
\be
v_{ij}=iG\left[(I_sI_fI_c)_i\,(I_sI_fI_c)_j-
(\gamma_5\mbox{\boldmath $\tau$}I_c)_i\cdot
(\gamma_5\mbox{\boldmath $\tau$}I_c)_j\right] \eqn{kij}
\ee
where $I_s, I_f\hspace{1mm}\mbox{and}\hspace{1mm}I_c$ are the unity operators in
the Dirac, flavour and colour spaces respectively, with the subscript $i$ ($j$)
indicating that the corresponding operators act in the $i$-th ($j$-th) quark's
one-particle space. In \eq{kij} and in the equations below we treat the quarks
as distinguishable particles as the inclusion of antisymmetrisation can always
be taken care of at the end. In the mentioned BS-like integral equation, the
quark propagator $d(p)$ satisfies the (nonlinear) Dyson-Schwinger equation
\be
d(p)=d_0(p)+d_0(p)\Sigma(p)d(p)    \eqn{DS}
\ee
where $d_0(p)$ is the bare quark propagator and the dressing term $\Sigma$ is
taken in the so-called Hartree approximation:
\be
\Sigma(p)=iG\int \frac{d^4k}{(2\pi)^4}\{\Lambda^\mu d(k)\Lambda_\mu
-\Lambda^\mu\mbox{tr}[d(k)\Lambda_\mu]\}.
\eqn{Hartree}
\ee
Here $\Lambda_\mu$ is the Lorentz four-vector $(I_sI_fI_c,
\gamma_5\mbox{\boldmath $\tau$}I_c)$, and the trace ``tr'' is over the Dirac,
flavour and colour indices.
%
The kernel of \eq{kij} is effectively separable so that the original three-body
equation for the three-quark system can be reduced down to a quark-diquark
two-body equation, and it is this latter form which we shall use to calculate
the electromagnetic corrections. In the channel where the two
interacting quarks form a scalar, isoscalar, positive parity diquark with colour
$\bar 3$, the kernel of \eq{kij} reduces to
\be
v_{f_1f_2,i_1i_2}=4ig_s(\gamma_5C\tau_2\beta^a)_{f_1f_2}\times
(C^{-1}\gamma_5\tau_2\beta^a)_{i_1i_2};\hspace{5mm}\beta^a_{ik}=
i\sqrt{\frac{3}{2}}\epsilon_{aik},\hspace{5mm}C=i\gamma_2\gamma_0
\eqn{scalar}
\ee
where $i_1i_2$ $(f_1f_2)$ are triples of initial (final) quantum numbers of the
first and second particle \cite{njl}. Then the diquark propagator is
\be
D(p)=\frac{4ig_s}{1-2ig_s\Pi(p^2)}\hspace{5mm}\mbox{where}\hspace{5mm}
\Pi(p^2)\delta_{ij}=-\int
\frac{d^4k}{(2\pi)^4}\mbox{tr}\left[\gamma_5\tau_i d(p+k)\gamma_5\tau_j d(k)
\right] \eqn{D}
\ee
and the quark-diquark interaction kernel is given by the quark exchange term
\be
K(p',p)=\gamma_5 d(p'+p)\gamma_5\beta^a\beta^{a'}.  \eqn{Z}
\ee
With $G_0=dD$ in \eq{G}, the resulting equation for the quark-diquark Green
function has the diagrammatic form illustrated in \fig{bs}.
\begin{figure}[t]
\centerline{\epsfxsize=9cm\epsfbox{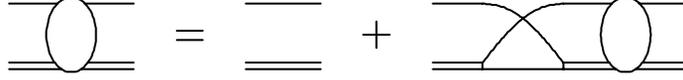}}
\vspace{3mm}

\caption{\fign{bs} The Bethe-Salpeter equation for the quark-diquark Green
function.}
\end{figure}
\section{ELECTROMAGNETIC CORRECTIONS TO THE NJL MODEL}

All electromagnetic corrections can be found by applying the general formulation
of \eqs{G}-(\ref{Delta}) to the particular case under consideration. Thus in the
case of a single quark, even though the solution of \eqs{DS} and
(\ref{Hartree}) is known to be
$d(p)=i(\pslash-m)^{-1}$, where $m$ is a constituent quark mass (which is not
zero even if $m_0=0$), these equations are needed for the proper construction of
the external and internal (with respect to the quark propagator) photon
currents. By identifying \eq{DS} as a special non-linear case of \eq{G}
(non-linear because $\Sigma$ contains $d$) and gauging both \eq{DS} and
\eq{Hartree}, we obtain that the electromagnetic current $d^\mu$ of the dressed
quark propagator satisfies the equation
\be
d^\mu(p)=d(p+q)\left[\gamma^\mu
+iG\int \frac{d^4k}{(2\pi)^4}\{\Lambda^\alpha d^\mu(k)\Lambda_\alpha
-\Lambda^\alpha\mbox{tr}[d^\mu(k)\Lambda_\alpha]\}\right] d(p) \eqn{d^mu}
\ee
which is linear in $d^\mu$ and can be easily solved [in \eq{d^mu} the
momentum of the incoming photon, $q$, is contained implicitly in all the $d^\mu$
functions].
The electromagnetic corrections to the dressed quark propagator, $\delta d(p)$,
can then be found in a similar manner.

We can similarly identify \eq{D} for the diquark propagator $D$ with \eq{G} and
therefore write down the diquark current $D^\mu$ and the electromagnetic
corrections to the diquark propagator $\delta D$ as
\be
D^\mu=D\Pi^\mu D;\hspace{2cm}\delta D=D\delta \Pi D+
D\Pi^\mu D\Pi^\nu DD_{\mu\nu}.
\ee

Finally, we can write down the complete set of electromagnetic corrections
$\delta G$ corresponding to the three-quark NJL model by identifying the
quark-diquark equation of \fig{bs} with \eq{G}. The resulting expression
for $\Delta$ is shown diagrammatically in \fig{fig:Delta}.
\begin{figure}[t]
\centerline{\epsfxsize=16cm\epsfbox{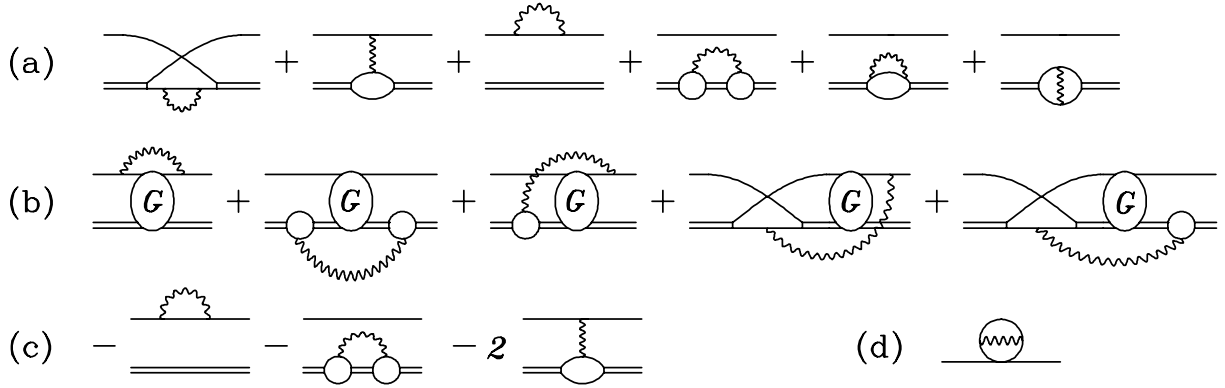}}
\vspace{4mm}

\caption{\fign{fig:Delta} The electromagnetic corrections to the NJL
model. Shown are the diagrams making up the $\Delta$ of \eq{Delta}: (a) $\delta
K$ (first diagram) and $G_0^{-1}\delta G_0 G_0^{-1}$ (the rest), (b) diagrams of
$\Gamma^\mu G\Gamma^\nu D_{\mu\nu}$ (also contributing are the last three
diagrams with initial and final states interchanged), and (c) diagrams making up
the subtraction term $-\Gamma_0^\mu G_0\Gamma_0^\nu D_{\mu\nu}$. For lack of
space we have not shown the corrections to the quark propagators in (a) due to
diagrams such as (d).
}
\end{figure}

Clearly, our procedure for finding a complete set of electromagnetic
corrections in the NJL model is of a general nature, and can be
used to include complete sets of lowest order corrections due to other
particle exchanges. Indeed we have recently applied our method to derive the
complete set of pionic corrections to the three-quark NJL model (where
previously only a part of such corrections were included \cite{ishii}), as well
as to some constituent quark models with confinement \cite{KB} (thereby
clarifying some recent discussions regarding this matter \cite{TG,Glozman}).
As our pionic corrections are complete, axial current is conserved exactly in
the limit of massless bare quarks. This feature is important for maintaining
chiral symmetry in the next to leading order approximations.

\end{document}